# $Tl_2LaCl_5$ ($Ce^{3+}$): New Fast and Efficient Scintillator for X- and γ-ray detection

H. J. Kim,[1,a)] Gul Rooh[2)] and Sunghwan Kim[3)]

[1]*Department of Physics, Kyungpook National University, Daegu 41566, Korea*

[2]*Department of Physics, Abdul Wali Khan University, Mardan, 23200, Pakistan*

[3]*Department of Radiological Science, Cheongju University, Cheongju 41566, Korea*

Scintillation properties of the new $Tl_2LaCl_5$: $xCe^{3+}$ where x = 0, 0.5, 1 and 10 mole % (TLC: $Ce^{3+}$) single crystals are presented. Two zones vertical Bridgman technique is used for the growth of this scintillation material. High Z-number (79) of this material offer excellent detection efficiency for X- and γ-rays. Grown samples are characterized under X- and γ-rays excitation in order to find the emission wavelength, energy resolution, light yield, and decay time. Emission spectra exhibit $Ce^{3+}$ emission bands between 350 - 550 nm and peaking at 389 nm. The obtained energy resolution and light yield increases with high Ce-concentration. Energy resolution and light yield of 10%Ce-doped sample are found to be 6.9% (FWHM) and 51,000±5000 ph/MeV, respectively at room temperature. For the grown samples, two exponential decay time constants are found. The measured decay time constants showed variation in their values with respect to Ce-concentrations in the host crystal. Fast decay time constant of 31 ns with 87% light emission is found for 10%Ce sample. Scintillation results suggested that TLC will be one of the promising next generation scintillator for the medical imaging techniques such as Positron Emission Tomography (PET).

## INTRODUCTION

Scintillation material has the ability to convert the ionizing radiations i.e. X- or γ- ray into visible or ultraviolet light. Since the discovery of scintillator it belongs to most important nuclear radiation detectors. Scintillation detectors are widely used in different fields which include medical diagnostic, experimental nuclear physics, high-energy physics, environmental studies, geological exploration, many fields of Physics and Chemistry[1-5]. For medical imaging techniques such as positron emission tomography (PET), desirable properties of a scintillator are high effective Z-number, high density, light yield and fast scintillation decay time. In commercial whole-body PET scanners variety of scintillators is used which includes, $Bi_4Ge_3O_{12}$ (BGO) [5-8], $Lu_2SiO_5$: Ce (LSO) [4, 9, 10], $Gd_2SiO_5$: Ce (GSO) [11], $Lu_{0.6}Y_{1.4}SiO_{0.5}$: Ce (LYSO) [12]. The performance of these scintillators is limited due to lacking of some scintillation properties. For example, LSO and LYSO are naturally radioactive due to $^{176}$Lu radioactive isotope which causes an unavoidable background level. Similarly, NaI: Tl, BGO and GSO has low light yield and slow decay time. Therefore, it is most important to explore new scintillation compounds in order to achieve the desirable scintillation properties for such application.

___

[a)] Corresponding author: Tel.:+82-53-950-5323; fax: +82-53-956-1739. E-mail address: hongjoo@knu.ac.kr (H.J. Kim).

Very recently, thallium based inorganic halide scintillators activated by cerium ion are reported by our group [13-15], these scintillators showed excellent scintillation performance. Due to high Z-number of thallium ion, these scintillators will offer high detection efficiency for X- and γ-ray in different applications. Moreover, the densities of the reported scintillators are comparable with NaI: Tl and CsI: Tl scintillators and could replace conventional scintillators in different applications. In this study, we present the growth and scintillation properties of the newly grown crystal, $Tl_2LaCl_5$: $xCe^{3+}$ where x = 0, 0.5, 1 and 10 mole % (TLC). Two zones vertical Bridgman technique is used for the growth of this scintillator. The scintillation properties of the grown samples include the measurements of the X-ray excitation emission spectra, decay time, energy resolution, and scintillation light yield.

## EXPERIMENTAL SECTION

### Crystal growth

Cerium doped single crystals of TLC has been grown by two zone vertical Bridgman technique. Stoichiometric amounts of TlCl (99.999%, Alfa-Aesar), $LaCl_3$ (99.999%, Sigma-Aldrich) and $CeCl_3$ (99.999%, Sigma-Aldrich) powders were weighing and loaded in quartz ampoule inside argon purged glovebox. All the loaded ampoules were sealed under high vacuum and were grown by two zone vertical Bridgman technique. Whole material was found to be melted at 510 °C. Sample crystals were transparent and homogenous but few cracks were appeared in the body. However, crack free samples with the dimensions of Ø 8 x 2 $mm^3$ were cut from the ingot for characterizations. Figure 1 shows the as grown sample of TLC single crystal. Due to hygroscopic nature all the grown samples were kept in mineral oil to avoid the degradation.

### Measurements of Emission Spectra and Scintillation Characteristics

For the measurement of X-ray induced luminescence spectra of TLC: $Ce^{3+}$ crystals, an X-ray tube having a W anode from a DRGEM. Co was used at room temperature. The X-ray generator was operated at power settings of 100 kV and 1 mA. A QE65000 fiber optic spectrometer made by Ocean Optics was used to measure the X-ray excitation spectra of the sample crystals. In order to measure the pulse height spectra of the TLC: $Ce^{3+}$ crystals, all the grown samples were wrapped in several layers of Teflon tape with one face left uncovered and directly coupled with the entrance window of the photomultiplier tube (PMT) (R6233, Hamamatsu) using index matching optical grease. For the signal generation in the detector, 662 keV γ-rays from $^{137}Cs$ source was used. The PMT signal were shaped with a Tennelec TC 245 spectroscopy amplifier and fed into a 25-MHz flash analog-to-digital converter (FADC) [16]. A software threshold was set to trigger an event by using a self-trigger algorithm on the field programmable gate array (FPGA) chip of the FADC board. The FADC output was recorded into a personal computer by using a USB2 connection, and the recorded data were analyzed with a C++ data analysis program [17].

Light yield of TLC crystals were measured at room temperature. A calibrated LYSO: $Ce^{3+}$ which has a calculated light yield 33,000 ph/MeV was used for the light yield measurement of TLC crystals. For the light yield measurement, similar experimental set up of pulse height measurement was used. Both crystals were directly attached with the entrance window of the (PMT) (R6233, Hamamatsu) were irradiated with $^{137}$Cs γ-ray source. Under similar conditions of PMT bias and amplifier gain, we recorded the pulse height spectra of the TLC and LYSO crystals. Based on the recorded pulse height spectra channel numbers of TLC and LYSO, the light yield of the grown samples of TLC were measured [18].

Decay time measurements of the TLC crystals were evaluated by using PMT (R6233, Hamamatsu) at a room temperature. After irradiation of TLC crystals with $^{137}$Cs γ-ray source, the signals generated in the PMT were fed into a 400 MHz FADC. A homemade FADC module was designed to sample the pulse every 2.5 ns for duration up to 64 μs so that to fully reconstruct each photoelectron pulse [19]. A trigger was formed in the FPGA chip on the FADC board. For low energy events, more than four photoelectrons in 2 μs were needed for the event trigger. An additional trigger was generated if the width of the pulse was longer than 200 ns for high energy events where many single photon signals were merged into a big pulse. The FADC located in a VME crate was read out by a Linux-operating PC through the VME-USB2 interface with a maximum data transfer rate of 10 Mbytes/s. The PMT output signals were fed into 400 MHz FADC and calculated the decay time spectra of the crystals using recorded pulse shape information [19].

## RESULTS AND DISCUSSION

### Crystal analysis

TLC single crystallized in Orthorhombic structure [20]. It has *Pnma* space group with lattice parameters of a = 12.833Å, b = 8.9750 Å and c = 8.0990 Å. Based on the lattice constants, the volume and density of the unit cell are calculated as 932.8 Å$^3$ and 5.20 g/cm$^3$, respectively. Effective Z-number of TLC is found to be 79, Table I presents comparison of the different properties of the TLC with other scintillation materials. Comparing with other scintillators see Table I, TLC will be a best choice for the detection of X- and γ-rays in many applications.

### X-ray induced luminescence

Figure 2 shows X-ray induced luminescence spectra of TLC: $Ce^{3+}$ doped crystals. Un-doped sample shows luminescence between 350 and 600 nm wavelength peaking around 400 nm. The origin of this emission is unknown. With increasing Ce-concentration in the host lattice, $Ce^{3+}$ like emission is observed (due to the lowest 5d excited state transition to the two spin–orbit split $^2F_{5/2}$ and $^2F_{7/2}$ ground state levels) which has almost similar emission peaks at 400 nm. Moreover, the broadness and intensities of the emission bands in the pure and 0.5% $Ce^{3+}$ spectra are disappeared with increase of Ce-concentrations i.e. 1% and 10% $Ce^{3+}$. This is might be due to the energy transfer mechanism between the host and $Ce^{3+}$-ions [21].

**Pulse height spectra and light yield**

Pulse height spectra of the TLC: $Ce^{3+}$ crystals are recorded under 662 keV γ-rays from $^{137}Cs$ source and shown in Fig. 3. The pulse height spectra reveals that the increase of $Ce^{3+}$ content in the host lattice improve the energy resolution of TLC crystals. From the pulse height spectra, we achieved best energy resolution of 6.9% (FWHM) for 10% $Ce^{3+}$ doped sample. Un-dope and 0.5% $Ce^{3+}$ samples shows 11.7% and 10.3% (FWHM) energy resolutions, respectively. Among the grown samples of TLC, 10% $Ce^{3+}$ doped sample showed highest light yield of 51,000±5000 ph/MeV under γ-ray excitation of 662 keV from a $^{137}Cs$ source at room temperature. Detailed results of the light yield of TLC: Ce crystals measured under γ-ray excitation at different shaping time of spectroscopic amplifier are listed in Table I. Figure 4 shows the comparison of the pulse height spectra of TLC: 10%$Ce^{3+}$ and LYSO crystals of the light yield measurement. In the Fig. 4, the photopeak position is proportional to the light yield of TLC: 10% $Ce^{3+}$ crystal. It is evident that the light yield of the TLC crystals depends on the Ce-concentration and overall 73% of light yield is increased from un-dope to 10% Ce-concentration. Therefore, we expect that further improvement is possible in light yield and energy resolution with the optimized condition of Ce-concentration of TLC crystal.

**Scintillation decay time**

Figure 6 shows the decay time spectra of TLC: $Ce^{3+}$ crystals under γ-ray excitation at room temperature. The decay time curves are fitted with the sum of two exponential functions and the obtained results are mentioned in Table II. From Table II, it is clear that the decay components of TLC crystals are get faster with the increase of Ce-concentration. 10% Ce doped sample shows fast decay component of 31 ns with 87% contribution of the total light yield. This implies that lager part of the total light yield of TLC: 10% $Ce^{3+}$ is emitted in the fast decay component. Moreover, the fast decay component is attributed to the life time of the 5d excited state of $Ce^{3+}$ ion [22]. Fast scintillators are required in different applications such as medical imaging, therefore it is expected that this scintillator could be used in the field of medical imaging.

## Conclusion

New single crystals of TLC compound activated with different Ce-concentrations are presented. For the single crystal growth of this material two zones vertical Bridgman technique are employed. The grown material have high Z-number and density and therefore offer higher detection efficiency of X-and γ-rays. TLC has a higher Z-number than the best available commercial scintillators with comparable density. Emission spectra under X-ray excitation shows $Ce^{3+}$ emission peaking at 400 nm. Among the measured scintillation properties, 10%Ce doped sample showed excellent performance. For the same dopant, an energy resolution of 6.9% (FWHM) and light yield of 51,000±5000 ph/MeV are obtained at under γ-ray excitation. Fast decay component of 31 ns with 87% total light yield contribution is obtained when10%Ce doped sample is excited with γ-rays. This decay component is attributed to the life time of 5d excited state of $Ce^{3+}$ ion. Decay components get faster with the increase of Ce-concentration in the host lattice. Similar trend is also observed in energy resolution and light yield, therefore it is expected that further improvement in the scintillation properties is possible with higher Ce-concentrations. Due to good scintillation performance (high Z-number, good density, high light yield and fast scintillation response), TLC in principle is an attractive candidate for use in positron emission tomography. Improvement in the scintillation properties are observed for this material which will be reported in the future report.

## Acknowledgment

These investigations have been supported by the National Research Foundation of Korea (NRF) funded by the Korean government (MEST) (No2015R1A2A1A13001843.).

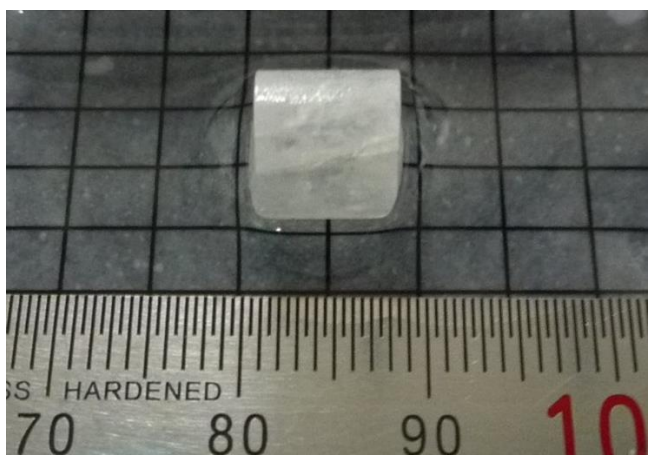

FIG. 1. Photograph of the grown TLC single crystal.

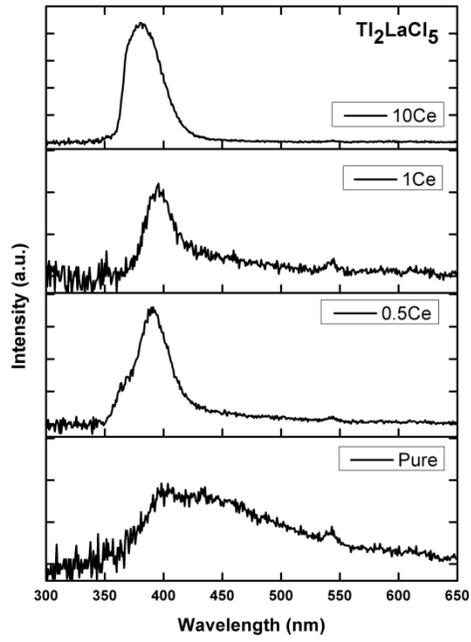

FIG. 2. Measured X-rays excited emission spectra of TLC single crystals for different Ce-concentrations.

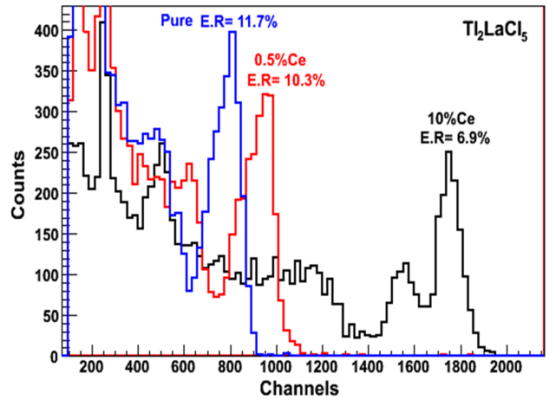

FIG. 3. Pulse height spectra of 662 keV γ-rays from a $^{137}$Cs source measured with TLC: $Ce^{3+}$.

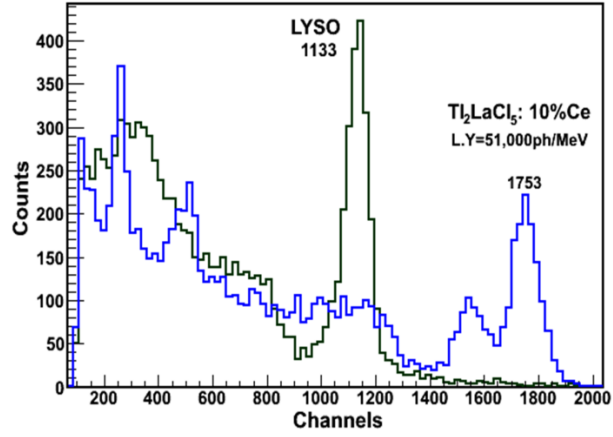

FIG. 4. Pulse height spectra of TLC: 10% $Ce^{3+}$ and LYSO: $Ce^{3+}$ crystals excited with γ-rays from a $^{137}Cs$ source. The photopeak channel numbers are proportional to the light yield.

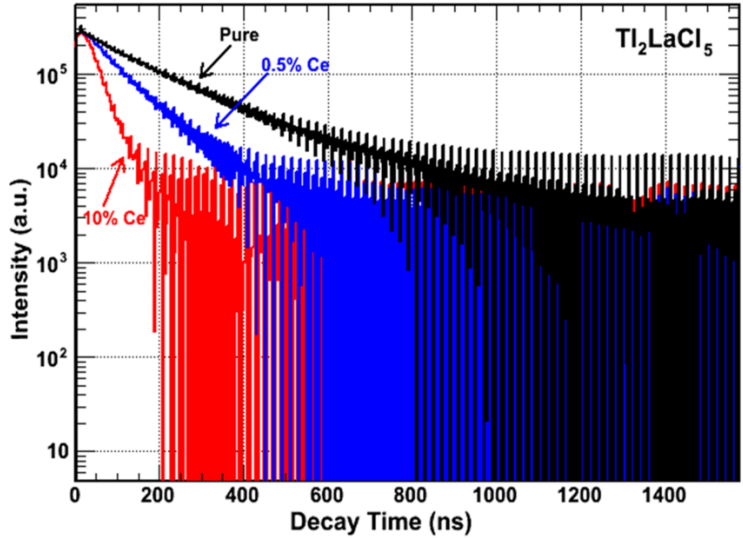

FIG. 5. Decay of $Ce^{3+}$ in TLC single crystals at room temperature under $^{137}Cs$ γ-ray excitation.

TABLE I. Comparison of TLC: $Ce^{3+}$ with different scintillators use in the PET application.

| Compound | $Z_{eff}$ | ρ (g/cm³) | $\lambda_{emission}$ (nm) | Light Yield (ph/MeV) at 662 keV | Energy resolution at 662 keV | Decay time (ns) |
|---|---|---|---|---|---|---|
| $Tl_2LaCl_5$: $Ce^{3+}$ | 79 | 5.2 | 400 | 51,000 | 6.9% | 31 |
| BGO: $Ce^{3+}$ [5, 12] | 73 | 7.1 | 480 | 8,000 | 15% | 300 |
| GSO: $Ce^{3+}$ [1,12] | 58 | 6.7 | 430 | 12,500 | 9.2% | 60 |
| LSO: $Ce^{3+}$ [1, 12, 23] | 66 | 7.4 | 420 | 27,000 | 12% | 40 |
| LYSO: $Ce^{3+}$ [1, 12, 23] | 65 | 7.1 | 420 | 32,000 | 8.7% | 53 |

TABLE II. Scintillation characteristics of TLC: Ce$^{3+}$ crystals at room temperature.

| Ce-concentration (mole%) | Energy resolution @ 662 keV(%) | Scintillation Light Yield (photons/MeV) | | | Decay time (ns) |
| --- | --- | --- | --- | --- | --- |
| | | 2 μs | 3μs | 6μs | |
| Pure | 11.7 | 27,000±2700 | 28,000±2800 | 29,500±2900 | 131ns (39%), 310ns (61%) |
| 0.5 | 10.3 | 38,000 ±3800 | 38,000 ±3800 | 38,000 ±3800 | 73ns (57%), 169ns (43%) |
| 10 | 6.9 | 49,000±4900 | 50,000±5000 | 51,000±5000 | 31ns (87%), 111ns (13%) |